\documentclass[12pt]{iopart}
\usepackage{cite}
\usepackage{graphics,graphicx}
\usepackage[hypertex]{hyperref}
\usepackage{iopams}
\usepackage{url}
\usepackage{soul}
\usepackage{xcolor}

\newcommand{\npone}{\mbox{$N\!+\!1$}}
\newcommand{\ao}{\mbox{$a_0\;$}}
\newcommand{\red}[1]{{\color{red} #1}}

\begin{document}


\title[e--OH collision cross sections]
{Calculated cross sections for low energy electron collision with OH}

\author{ K. Chakrabarti$^\ast$, V. Laporta$^\sharp$, and 
Jonathan Tennyson$^{\dag}$}
\address{
$^\ast$Department of Mathematics, Scottish Church College, 1 \&  3
Urquhart Sq., Kolkata 700006,  India \\
$^\sharp$Istituto per la Scienza e Tecnologia dei Plasmi, Consiglio Nazionale delle Ricerche, Via Amendola 122/D, Bari, 70126, Italy\\
$^\dag$Department of Physics and Astronomy, University College
London, Gower St., London WC1E 6BT, UK}

\ead{$^1$j.tennyson@ucl.ac.uk}
\begin{abstract} 
  The hydroxyl radical, OH, is an important component of many natural and technological plasmas
  but there is little available information on processes involving its
  collisions with low-energy electrons. Low-energy electron collisions
  with OH are studied in the framework of the R-matrix method. Potential
  energy curves of some of the low lying target states of doublet and
  quartet symmetry which go to the O($^3$P)+H($^2$S),
  O($^1$D)+H($^2$S) and O($^1$S)+H($^2$S) asymptotic limits are
  obtained for inter nuclear separations between $1-6~a_0$.
  Scattering calculations are performed at the OH equilibrium geometry
  $R_e=1.8342~a_0$ to yield cross sections for elastic scattering,
  electronic excitations form the $\mathrm{X}\,^2\Pi$ ground state to the first
  three excited states of $\mathrm{A}\,^2\Sigma^+$, $a\,^4\Sigma^-$, $1\,^2\Sigma^-$ 
	symmetry and for electron impact dissociation of OH. The positions 
	and widths for negative ion resonances in the $e$--OH system are used
	estimate the cross section for dissociative electron attachment to 
	OH which is found to be significant at electron energies about 1.5 eV.

\end{abstract}
\submitto{\PSST}
\maketitle

\section{Introduction}

The hydroxyl radical, OH, is key component of many plasmas include
atmospheric ones, particularly if the air is humid
\cite{16Bened,17Takeda,17yue,17Winters,18Schr,19Wang1,19Shahm} or near
liquid water \cite{16Li,17Hsieh,18Vorac}, and plasmas formed during
combustion \cite{17Ehn,19Wang2}. Models of such plasmas require rates
or cross sections for key processes occuring in the plasma but for OH
these are largely unknown \cite{16CaBr} and hence are absent from
major data compilations \cite{jt647,jt679}. Indeed the best source of
electron--OH rates \cite{18Schr} appears to be the weighted total
cross-section (WTCS) caclulations of Riahi {\it et al.} \cite{Riahi06}
which, for OH, are based on old and unproven electron collision cross sections
\cite{67Drawi}.

Given the difficulty in measuring electron collision
cross sections with open shell species such as OH, it would
appear to be timely to use theoretical methods to 
establish a reliable dataset of cross sections and rates
for
electron collisions with OH is expected to govern its chemistry 
and hence it is necessary to have reliable cross sections for 
different electron induced processes in OH.

The starting point of our collision calculation requires accurate 
potential energy curves (PECs) for OH. Several studies on the PECs 
of OH exist. Calculations on the OH doublet and quartet states were
performed as early as 1973 by Easson and Pryce \cite{Easson73}. 
Langhoff \etal \cite{Langhoff82} studied the ground and some of the 
$^2\Sigma^+$ states of OH 
to obtain their dipole moments and radiative lifetimes.
A configuration interaction (CI) study on the $^2\Sigma^-$ states
of OH was performed by van Dishoeck \etal \cite{vanDish83}. Subsequently, 
they extended this work to a systematic study of the ground (X~$^2\Pi$) and 
several excited states of doublet and quartet symmetry 
\cite{vanDish83b}. There are also some recent and more sophisticated
calculations using large GTO basis sets performed on some of the low 
lying OH states \cite{Srivastava14,Qin14}.

Compared to OH, its anionic states are much less studied even though
these are known to play an important role in collision processes such
as the dissociative electron attachment (DEA). Singlet and triplet
OH$^-$ states of $\Sigma^+$ and $\Pi$ symmetries were obtained 
by Sun and Freed \cite{Sun82} at the OH equilibrium bond length of
$R_e=1.8342$ \ao to calculate vertical excitation energies of the
excited states of OH$^-$. Several calculations on the PECs
of these singlet and triplet states of OH$^-$ also exist 
\cite{Komiha91,Chen97,Nemukhin97,Tell90}.
More recently, calculations on the OH$^-$ anionic states 
were performed with large GTO bases by 
Srivastava and Satyamurty \cite{Srivastava14} and Vamhindi \etal
\cite{Vamhindi16}. All of the studies indicate that the X~$^1\Sigma^+$ 
state of OH$^-$ is strongly bound where as the $^1\Pi$ and 
$^3\Pi$ excited states are only quasi-bound as these resonance
states lie in the  continuum of the OH neutral plus free electron system.

Suprisingly few studies on electron plus OH collisions exist.  A
limited study of the vibrationally inelastic cross sections, both
integrated and differential, for the excitation of the $\nu=1$
vibrational level of the electronic ground state was performed by Chen
and Morgan \cite{Chen97} in the energy range $0-3$ eV using the
R-matrix method.  A much more detailed calculation of the elastic
differential, integral, momentum-transfer cross sections as well as
grand-total (elastic + inelastic) and total absorption cross sections
for electron-OH collisions were performed by Sobrinho \etal
\cite{Sobrinho04} using the Schwinger variational method and the
distorted-wave approximation.  Since no other results were available
on \red{$e$}--OH collisions, Sobrinho \etal compared their cross sections with
the corresponding cross sections for \red{$e$}--H$_2$O collisions which they
found to be remarkably similar.  The ionisation potential for OH is 13 eV \cite{IP}.
As mentioned above, cross section for electron impact
ionisation and rate coefficients for OH(X~$^2\Pi$)$\rightarrow$
OH(A~$^2\Sigma^+$) excitations were obtained by Riahi \etal
\cite{Riahi06} using the WTCS theory which is
essentially a model calculation.

In the present work, we present cross sections and associated rates for the OH molecule for elastic scattering,
electron impact electronic excitation, 
\begin{eqnarray}
e + \rm{OH}(X\,^2\Pi) &\rightarrow & e + \rm{OH}(A\,^2\Pi, a\,^4\Sigma^-, 1\,^2\Sigma^-)\,,
\end{eqnarray}
and electron impact dissociation,
\begin{eqnarray}
e + \rm{OH}(X\,^2\Pi) &\rightarrow & e + \rm{O}(^3P) + \rm{H}(^2S)\,,
\\
&\rightarrow & e + \rm{O}(^1D) + \rm{H}(^2S)\,.
\end{eqnarray}

We also study electronic states of the OH$^-$ negative
ion and resonances in the $e$--OH system at the OH equilibrium
bond length $R_e=1.8342$ \ao. Finally, an estimate of the DEA cross section,
\begin{equation}
e^- + \rm{OH}(X\,^2\Pi) \rightarrow \rm{O}^-(^2P) + \rm{H}(^2S)\,,
\end{equation}
is made using a model proposed by Munro \etal \cite{Munro12};
we are not aware of the DEA of OH being included in models of OH-containing
plasmas.

The paper is organized as follow: Section \ref{sec:rmat_th} and Section 
\ref{sec:rmat_calc} report the theoretical framework of R-matrix and  the configurations for target and scattering
calculations respectively. Section \ref{sec:results}
shows the results for potential energy curves, couplings and cross sections. Finally, Section \ref{sec:conc} reports our conclusions and perspectives.

\section{The R-matrix Method}
\label{sec:rmat_th}
Our calculations are permormed using the {\it{R}}-matrix method the
details of which can be found in reviews by Burke \cite{Burke11} 
and Tennyson \cite{Tennyson10}. The R-matrix method  is based on division 
of the configuration space into an inner region, here a sphere 
of radius $11$ \ao centred at the molecular centre-of-mass, and an
outer region exterior to this sphere. In the inner region, the wave
function of the (\npone)-electron system (OH + $e^-$) is written as 
a close coupling (CC) expansion,
\begin{equation}\label{eq:cc}
\Psi_k = {\cal A} \sum_{i,j} a_{i,j,k} \Phi_i(1,\ldots,N)
F_{i,j}(\npone) + \sum_i b_{i,k} \chi_i(1,\ldots,\npone) \;,
\end{equation}
where $\cal A$ is the antisymmetrisation operator, $\Phi_i$ is the 
$N$-electron wave function of the $i^{th}$ target state, $F_{i,j}$ 
are continuum orbitals and $\chi_i$ are two-centre $L^2$ functions
constructed by making all $(N+1)$-electrons occupy the target molecular
orbitals (MOs), and takes into account the polarization of the 
$N$-electron target wave function in presence of the projectile 
electron.

An {\it{R}}-matrix is then built at the boundary of the {\it{R}}-
matrix sphere using the inner region wave function. The 
{\it{R}}-matrix is then propagated to asymptotic distances and 
matched with known asymptotic functions \cite{Noble84}. The
matching yields the $K$-matrix from which all scattering 
observables can be extracted.

As Slater type orbitals (STOs) are known to provide a better target
representation for diatomic targets, we used the diatomic version of
the UK molecular {\it{R}}-matrix codes \cite{Morgan98} which uses STOs
to represent the target wavefunction. The continuum was represented by
numerical orbitals in a partial wave expansion about the molecular
center of mass.  Since the OH target is neutral, the numerical orbitals
were chosen to be spherical Bessel functions. A Buttle correction
\cite{Buttle67} was also used to allow for the arbitrary fixed
boundary conditions imposed on the continuum basis orbitals.

\section{Calculations}
\label{sec:rmat_calc}
\subsection{The OH target}
A configuration interaction (CI) model was used to represent the OH
target.  As the choice of basis sets affect the quality of such
calculation we tested several STO basis sets, namely those of Cade and
Huo \cite{Cade67}, Emma \etal \cite{Emma03} and Langhoff \etal
\cite{Langhoff82}. Finally basis set II given in Table 1 of Langhoff
\etal was chosen as it gave excitation energies at the OH equilibrium
bond length $R_e=1.8342$~\ao that were in good agreement with other
calculations \cite{vanDish83b,Langhoff82, Srivastava14,Qin14}.
However, at larger bond lengths, the Langhoff \etal basis set gave
severe linear dependence in our calculation.  To mitigate this, one
$2s$ basis function centered on the H atom was deleted and the
resulting 'trimmed' basis set was used in all subsequent calculations.
This new basis set contained 10 $s$-type, 6 $p$-type, 2 $d$-type basis
functions centered on the O atom and 3 $s$-type and 2 $p$-type basis
functions centered on the H atom respectively.

The STOs were used to build a basis of 35 molecular orbitals 
consisting of $23\sigma$, $10\pi$ and $2\delta$ orbitals. An initial
set of SCF calculation was  made for the X~$^2\Pi$ and B~$^2\Sigma^-$ 
states of OH. Two sets of natural orbitals (NOs) of $^2\Pi$ and
$^2\Sigma^-$ symmetry were then obtained by doing a complete active 
space + singles and doubles (CAS+SD) calculation using these SCF 
orbitals. The NOs were then used in a subsequent CI calculation.

Tests showed that the target excitation energies were sensitive to 
the choice of the natural orbitals included in the CAS CI calculations. 
A compromise set was therefore chosen in which the $3\sigma,4\sigma$ 
and $5\sigma$ target orbitals were represented by $^2\Sigma^-$ NOs 
and the remaining $\sigma,\pi$ and $\delta$ target orbitals by 
$^2\Pi$ NOs. 

We tested several target models. Of these, the model denoted
$(1\sigma)^2$ $(2\sigma-6\sigma, 1\pi-2\pi)^7$ was selected; this model
has
the $1\sigma$ orbital frozen and the CAS was defined by 
$(2\sigma-6\sigma, 1\pi-2\pi)^7$. This model gave the best target
ground state and excitation energies.

\begin{figure}[h!]
\begin{center}
\includegraphics[scale=0.5,viewport=0 0 723 531, clip=true]{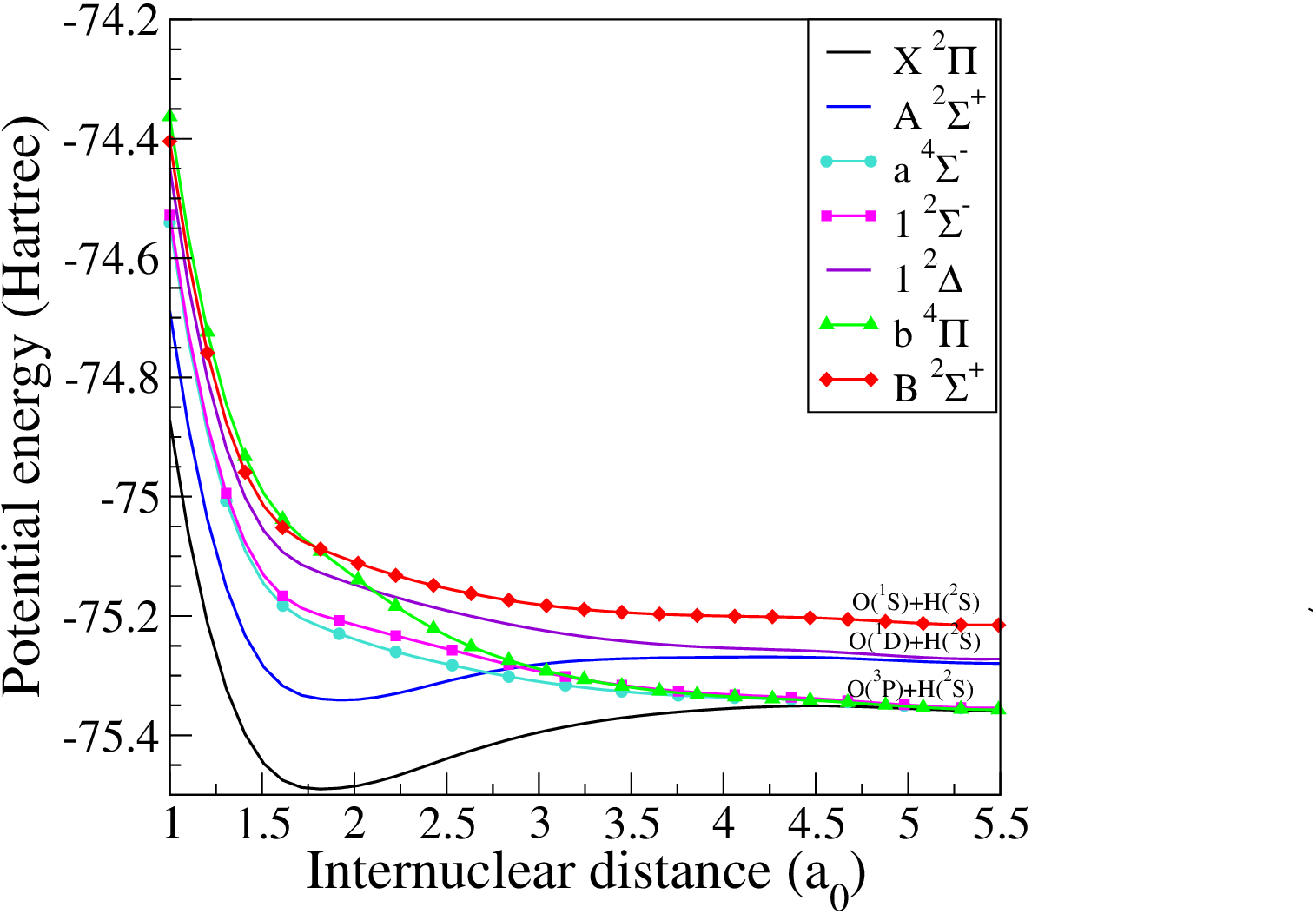}
\end{center}
\caption{Potential energy curves of the first eight OH target states.
\label{fig:pec}}
\end{figure}

Figure \ref{fig:pec} shows the behavior of eight lowest states
of OH used in our calculation. The asymptotic limits of each of these
curves are also shown in the figure. In common with previous electronic
structure studies \cite{Qin14}, we find that curves above the first
excited states are repulsive.

The vertical excitation energies of some of the doublet and quartet
states of OH at its equilibrium bond length, $R_e=1.8342$ \ao, are shown 
in Table \ref{tab:excitation}. These are compared with the calculations 
of van Dishoeck \etal \cite{vanDish83b} (which are more comprehensive)
and those in Refs. \cite{Langhoff82,Srivastava14,Qin14}. The excitation
energies are in good agreement with those of van Dishoeck \etal except 
for the a~$^4\Sigma^-$ and the $b~^4\Pi$ states which are estimated form 
the corresponding PECs. Note also that  van Dishoeck \etal used a different 
value, $R_e=1.85$ \ao, for their equilibrium bond length. The excitation
energy for the D~$^2\Sigma^+$ state appears to be higher than that
quoted in Qin and Zhang \cite{Qin14} as their value is adiabatic.

The dipole moment of the X~$^2\Pi$ state was found to be 1.602 D
and is in good agreement with the corresponding MCSCF value 1.612 D
of Werner \etal \cite{Werner83} and the measured value for the
vibrational ground state of 1.655 D \cite{84PeFrKl}.

\subsection{Scattering calculations}
Our calculations are performed at a single geometry, namely the OH
equilibrium geometry $R_e=1.8342$ \ao. We have used 14 OH natural orbitals
$(8\sigma, 4\pi, 2\delta)$ and a $(2-6\sigma, 1-2\pi)^7$ CAS target
wave functions which allows for 2 virtual orbitals per symmetry. These 
were augmented by continuum orbitals $F_{ij}$ expressed as a truncated
partial wave expansion about the center of mass retaining partial waves
with $l \le 6$ and $m \le 2$ in the expansion. Since the target was
neutral, the radial parts of the continuum function was chosen to be
spherical Bessel functions and solutions below 109 eV were retained. 
A Buttle \cite{Buttle67} correction was used to correct for the effect 
of fixed boundary condition used to generate the functions. This 
produced $59\sigma$, $49\pi$, $40\delta$ continuum functions which 
were then Schmidt orthogonalized to the target NOs.

Scattering calculations were performed on the $^1\Sigma^+$, $^1\Sigma^-$,
$^1\Pi$, $^1\Delta$, $^3\Sigma^+$, $^3\Sigma^-$, $^3\Pi$ and $^3\Delta$
total symmetries of the $e+$OH system. The summary of the target states 
used for each symmetry in the CC expansion Eq.~(\ref{eq:cc}) is shown in
Table \ref{tab:states}. Since the contribution to the  cross sections
from calculations of $\Delta$ symmetry  were
already found to be small, we did not consider  higher symmetries 
in the calculations.

\begin{table}[h!]
\caption{Vertical excitation energies (in eV) from the X\,$^2\Pi$
ground states of the OH molecule at OH equilibrium bond length
$R_e=1.8342$ \ao. The absolute energy of the X\,$^2\Pi$ ground state 
is $-75.490110$ Hartree. Also given are our computed
the absolute (transition) dipole moment ($\mu$).} 
\label{tab:excitation}
\begin{tabular} {lcccccc} 
\hline
OH state & This work & van Dishoeck \etal$^a$ & Langhoff \etal$^b$ & 
Ref. \cite{Srivastava14}$^\dag$ & Ref. \cite{Qin14}$^\dag$& $\mu$/D\\
\hline
X\,$^2\Pi$      & 0.0  & 0.0  & 0.0  & 0.0 & 0 & 1.60\\ 
A\,$^2\Sigma^+$ & 4.05 & 4.05 & 4.063 & 4.1 & 4.05& 0.63\\
a\,$^4\Sigma^-$ & 7.36 & 6.7* & -- & -- & --\\
1\,$^2\Sigma^-$ & 7.89 & 7.20 & -- & -- & --& 1.45\\
1\,$^2\Delta$   & 9.81 & 9.33 & -- & -- & --& 1.15\\ 
b\,$^4\Pi$      & 10.72 & 9.5* & -- & -- & --\\
B\,$^2\Sigma^+$ & 10.88 & 10.98 & 11.192 & -- & 8.75&0.94\\
\hline\\
\end{tabular}\\
$^a$ van Dishoeck \etal \cite{vanDish83b} ($R_e=1.85$ \ao)\\
$^b$ Langhoff \etal \cite{Langhoff82}\\
$^*$Estimated from the corresponding potential energy curve.\\
$^\dag$ Experimental adiabatic values.\\
\end{table}

\begin{table}
\caption{Symmetry and number of states used in the close-coupling expansion Eq.(\ref{eq:cc}).
The target states of lowest energy were used in each case.}
\label{tab:states}
\begin{tabular} {lcl} 
\hline\\
Symmetry & Number & Target states coupled\\
\hline\\
$^1\Sigma^+$ & 4 & one $^2\Pi$, two $^2\Sigma^+$ and one $^2\Delta$
 states\\
$^1\Sigma^-$ & 3 & one each of $^2\Pi$, $^2\Sigma^-$, $^2\Delta$
states\\
$^1\Pi$ & 4 & one each of $^2\Pi$, $^2\Sigma^+$, $^2\Sigma^-$ and $^2
\Delta$ states\\
$^1\Delta$ & 4 & one each of $^2\Pi$, $^2\Sigma^+$, $^2\Sigma^-$ and $^
2\Delta$ states\\ 
$^3\Sigma^+$ & 4 & one each of $^2\Pi$, $^2\Sigma^+$, $^2\Delta$ and $^
4\Pi$ states\\
$^3\Sigma^-$ & 5 & one each of $^2\Pi$, $^2\Sigma^-$, $^2\Delta$, $^4
\Pi$, and $^4\Sigma^-$ states\\
$^3\Pi$ & 5 & one each of $^2\Pi$, $^2\Sigma^+$, $^2\Sigma^-$, $^4\Pi$,
and $^4\Sigma^-$ states\\
$^3\Delta$ & 6 & one each of $^2\Pi$, $^2\Sigma^+$, $^2\Sigma^-$, $^2
\Delta$, $^4\Pi$, and $^4\Sigma^-$ states\\
\hline
\end{tabular}
\end{table}

\section{Results}
\label{sec:results}
In the following subsections, we present our results for the OH$^-$ bound 
states, resonance positions and widths at $R_e=1.8342$ \ao, cross sections 
for elastic scattering, electronic excitations and estimate of the DEA
cross sections. A more complete study of the DEA process would require
detailed negative ion resonance curves and widths as a function of 
geometry and we propose to undertake this in a subsequent paper. To the 
best of our knowledge, results for the cross sections presented here
have never been studied before.

\subsection{Bound and resonant states}
The inner region solutions obtained were used to construct an $R$-matrix
on the boundary. In the outer region, the potential was given by 
the diagonal and off-diagonal dipole moments of the OH target states. 
The $R$-matrices were propagated in this potential to 50 \ao and then matched
with exponentially decaying functions obtained from a Gailitis expansion
\cite{Noble84}. To find the bound states, the searching algorithm of 
Sarpal \etal \cite{Sarpal91} with the improved nonlinear, quantum defect 
based grid of Rabad\'an and Tennyson \cite{Rabadan96} was used. This 
method, originally developed by Seaton \cite{Seaton85}, searches for the 
zeros of an energy dependent determinant $\bold{B}(E)$ using either an 
energy or a quantum defect grid. The zeros of $\bold{B}(E)$ can be shown
to correspond to the bound state energies.

For resonance calculation, the $R$-matrix was propagated  to 70 \ao to
obtain stable results. It was then matched with Coulomb functions
using the Gailitis expansion procedure of Noble and Nesbet
\cite{Noble84}.  The recursive program RESON in the $R$-matrix code
suit \cite{Tennyson84} was used to detect resonances and fit the
resonances to a Breit-Wigner profile to obtain their energies ($E$)
and width $(\Gamma)$ with an initial energy grid $0.5\times 10^{-3}$ Ryd.

The resonance positions and widths of some of the low lying resonances
that are relevant for DEA are shown in Table \ref{tab:reson}. Interestingly,
the $1\,^3\Sigma^+$ resonance is given by  Vamhindi and Nsangou 
\cite{Vamhindi16} but not by Srivastava and Satyamurthy \cite{Srivastava14},
even though both these works study OH$^-$ resonant states in some detail. 

\begin{table}[h!]
\caption{Resonance positions and widths (in eV) of some of the low lying
Feshbach resonances in the $e$--OH at OH equilibrium bond length. Figures
in brackets indicate power of ten.
$R=1.8342$ \ao.} 
\label{tab:reson}
\begin{tabular} {lcc} 
\hline
State & Position & Width\\
\hline
Below A\,$^2\Sigma^+$ state &&\\
a\,$^3\Pi$      & 2.447 & 0.0315 \\
A\,$^1\Pi$      & 2.536  & 0.0441\\
Below a\,$^4\Sigma^-$ state &&\\
1\,$^3\Sigma^+$ & 6.403 & 0.0454 \\
\hline\\
\end{tabular}\\
\end{table}

A limited number of studies on the OH$^-$ states exist of which those of
Refs.~\cite{Sun82, Tell90, Chen97, Srivastava14, Vamhindi16} are noteworthy. 
In particular, Chen and Morgan \cite{Chen97}, Srivastava and Satyamurthy
\cite{Srivastava14} and Vamhindi and Nsangou \cite{Vamhindi16} obtained
potential energy curves for some of the OH$^-$ states. These studies 
indicate that only the X\,$^1\Sigma^+$ state of OH$^-$ is stable and bound. 
The a\,$^3\Pi$, A\,$^1\Pi$, $1\,^3\Sigma^+$ lie above the parent neutral
X\,$^2\Pi$ state for $R \le 3.0$ \ao and are of resonant character. 
However, for larger internuclear distances, these states become stable 
as they lie below the neutral X\,$^2\Pi$ state. The b\,$^3\Pi$ was obtained
only by Srivastava and Satyamurthy and is of fully resonant character
as it lies above the OH(X\,$^2\Pi$) ground state for internuclear distances $R$.

For a comparison of the relative positions of the $^1\Pi$ and $^3\Pi$
resonant states whose parent is OH(X\,$^2\Pi$), we computed the
vertical excitation energies of these states and compare them with available
results in Table \ref{tab:ve}. Except for the corresponding results of
Sun and Freed \cite{Sun82} which appear too high, all other results
are in good agreement with each other.

Table \ref{tab:ea} compares the energy difference $\Delta E$
between the OH(X\,$^2\Pi$) ground state and the X\,$^1\Sigma^+$, 
a\,$^3\Pi$ and A\,$^1\Pi$ states of OH. $\Delta E$ for the X\,$^1\Sigma^+$
state is the electron affinity, $E_a$, of OH. Our results are compared
with those of Srivastava and Sharma \cite{Srivastava14}, Chen and
Morgan \cite{Chen97}, Werner \etal \cite{Werner83} and the 
experimental value of $E_a$ given by Schulz \etal \cite{Schulz82}.
Our values of $\Delta E$ agree reasonably with those of 
Srivastava and Sharma. Our computed value of the electron affinity 
of OH is also in close agreement with all other results, and deviates 
by about 0.27 eV from the experiment value \cite{Schulz82}.

\begin{table}[h!]
\caption{Vertical excitation energies (in eV) from the X\,$^1\Sigma^+$
ground states of the OH$^-$ molecule at the OH equilibrium bond 
length $R_e=1.8342$ \ao. Our absolute energy of the X\,$^1\Sigma^+$ 
ground state is $-75.547590$ Hartree.} 
\label{tab:ve}
\begin{tabular} {lcccc} 
\hline
OH$^-$ state & This work & Srivastava \&\ Satyamurthy$^a$
&Sun \&\ Freed$^b$ & Tellinghuisen \etal\\

\hline
X\,$^1\Sigma^+$ & 0.0  & 0.0  & 0.0 & 0.0 \\ 
a\,$^3\Pi$ & 3.89 & 3.72 & 9.67 & 3.47\\
A\,$^1\Pi$ & 4.01 &  3.93 & 10.62 & 3.75 \\
\hline\\
\end{tabular}\\
$^a$ Srivastava and Satyamurthy \cite{Srivastava14}\\
$^b$ Sun and Freed \cite{Sun82}\\
$^c$ Tellinghuisen \etal \cite{Tell90}\\
\end{table}

\begin{table}[h!]
\caption{Energy difference $\Delta E$ (in eV) between the 
OH(X\,$^2\Pi$) ground state and the X\,$^1\Sigma^+$, a\,$^3\Pi$ 
and A\,$^1\Pi$ states of the OH$^-$ molecule at OH equilibrium 
bond length $R_e=1.8342$ \ao. $\Delta E$ for the X\,$^1\Sigma^+$ 
state is the electron affinity, $E_a$, of OH.} 
\label{tab:ea}
\begin{tabular} {lccccc} 
\hline
OH$^-$ state & This work & Srivastava \&\ Satyamurthy $^a$
&Chen \&\ Morgan$^b$ & Werner \etal$^c$ & Expt.$^d$\\

\hline
X\,$^1\Sigma^+$ & 1.56  & 1.87  & 2.14 & 1.58 & 1.83\\ 
A\,$^1\Pi$ & -2.53 & -2.06 & -- & -- & --\\
a\,$^3\Pi$ & -2.41 &  -1.85 & -- & -- & --\\
\hline\\
\end{tabular}\\
$^a$ Srivastava and Satyamurthy  \cite{Srivastava14}\\
$^b$ Chen and Morgan \cite{Chen97}\\
$^c$ Werner \etal \cite{Werner83}\\
$^d$ Schulz \etal \cite{Schulz82}\\
\end{table}

\begin{figure}[h!]
\begin{center}
\includegraphics[scale=0.5,viewport=0 0 723 375, clip=true]{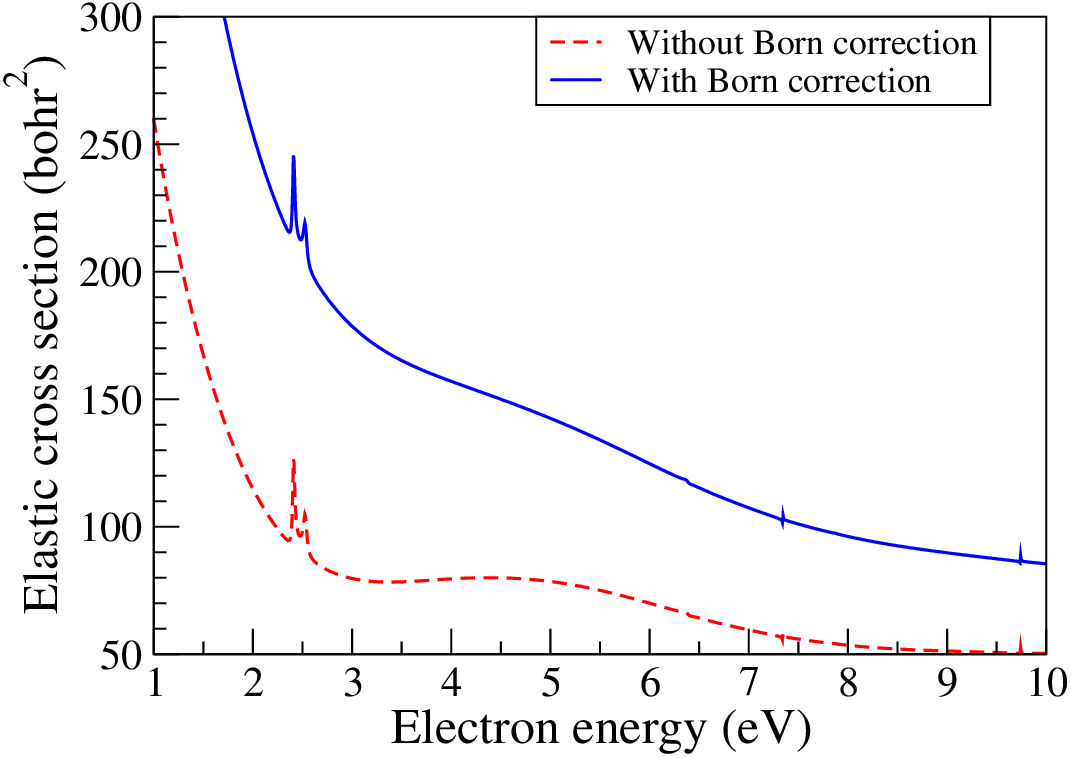}
\end{center}
\caption{Elastic cross section for electron impact on OH(X\,$^2\Pi$) ground state at $R_e=1.8342$ $a_0$. 
\label{Fig:Elastic}}
\end{figure}

\begin{figure}[h!]
\begin{center}
\includegraphics[scale=0.4,viewport=0 0 530 406, clip=true]{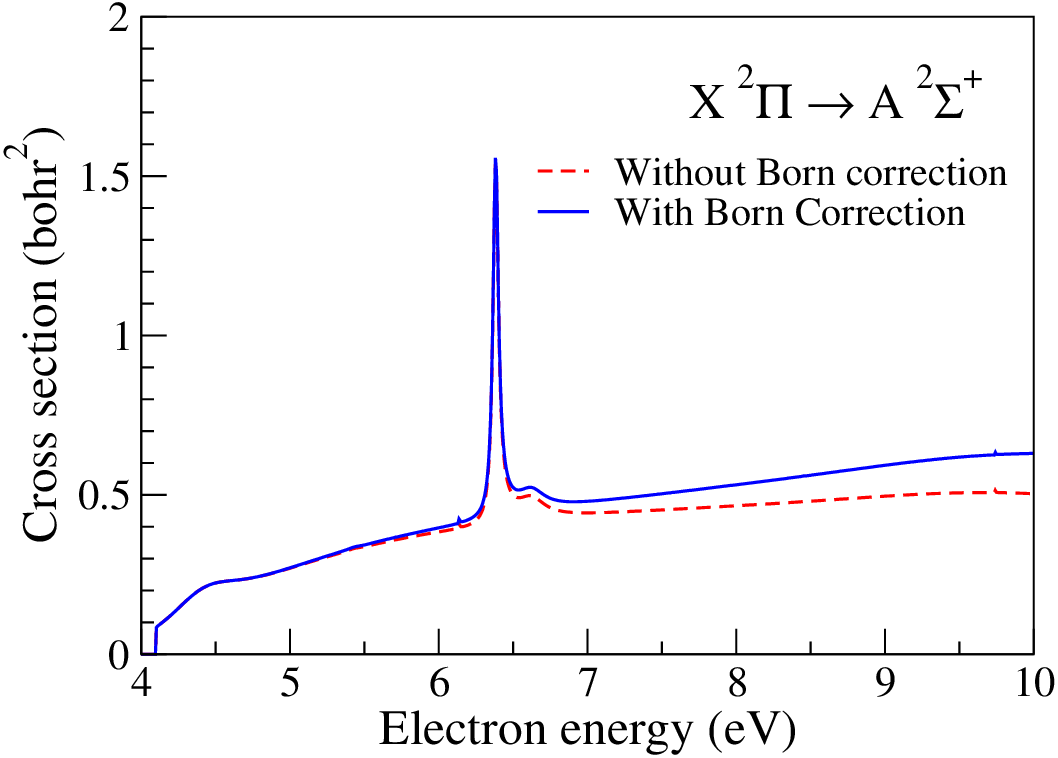}
\includegraphics[scale=0.4,viewport=0 0 528 406, clip=true]{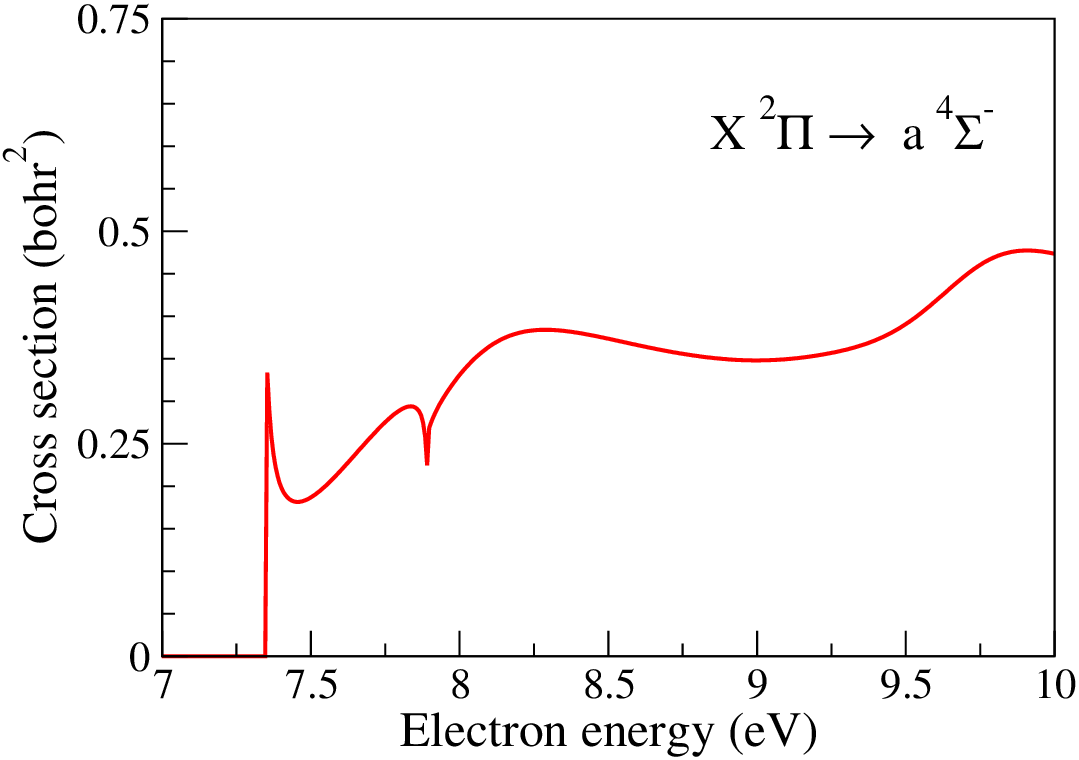}
\includegraphics[scale=0.4,viewport=0 0 531 400, clip=true]{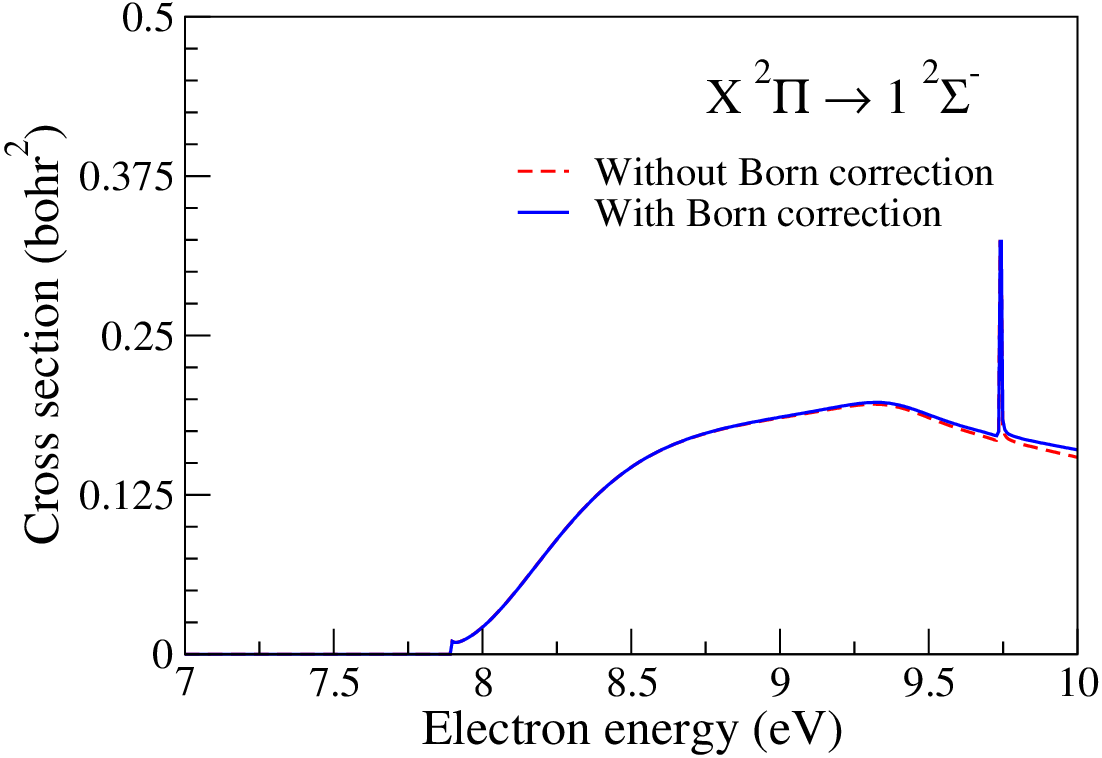}
\end{center}
\caption{Excitation cross sections from the X\,$^2 \Pi$ ground
state of the OH molecule to the excited states shown in each
panel for $R_e = 1.8342$ $a_0$.
\label{Fig:Excitation}}
\end{figure}

\subsection{Cross sections} 
\subsubsection{Elastic scattering, electronic excitation and
dissociation} 

Figure \ref{Fig:Elastic} shows cross sections for elastic scattering 
from OH. A Born correction to the cross sections was done to include
contribution from higher partial waves. As is usual for molecules
with a significant permanent dipole moment \cite{jt464}, the elastic cross section is 
strongly peaked at low energies due the  
strong forward scattering associated with the long-range potential due to
the dipole moment. This effect
is difficult to detect experimentally but has very recently been
observed in low-energy collisions with water using specially designed apparatus
sensitive to scattering angles of less than 3.5$^\circ$\cite{jt769}. The elastic cross section also features sharp peaks around 
2.5 eV due to capture into the OH$^-$ A\,$^1\Pi$ and a\,$^3\Pi$ 
resonant states. There is also a broad shape resonance like feature 
around 5 eV which may be due to temporary capture into resonant states 
with excited OH states as parent states.

\begin{figure}[h!]
\begin{center}
\includegraphics[scale=0.5,viewport=0 0 723 375, clip=true]{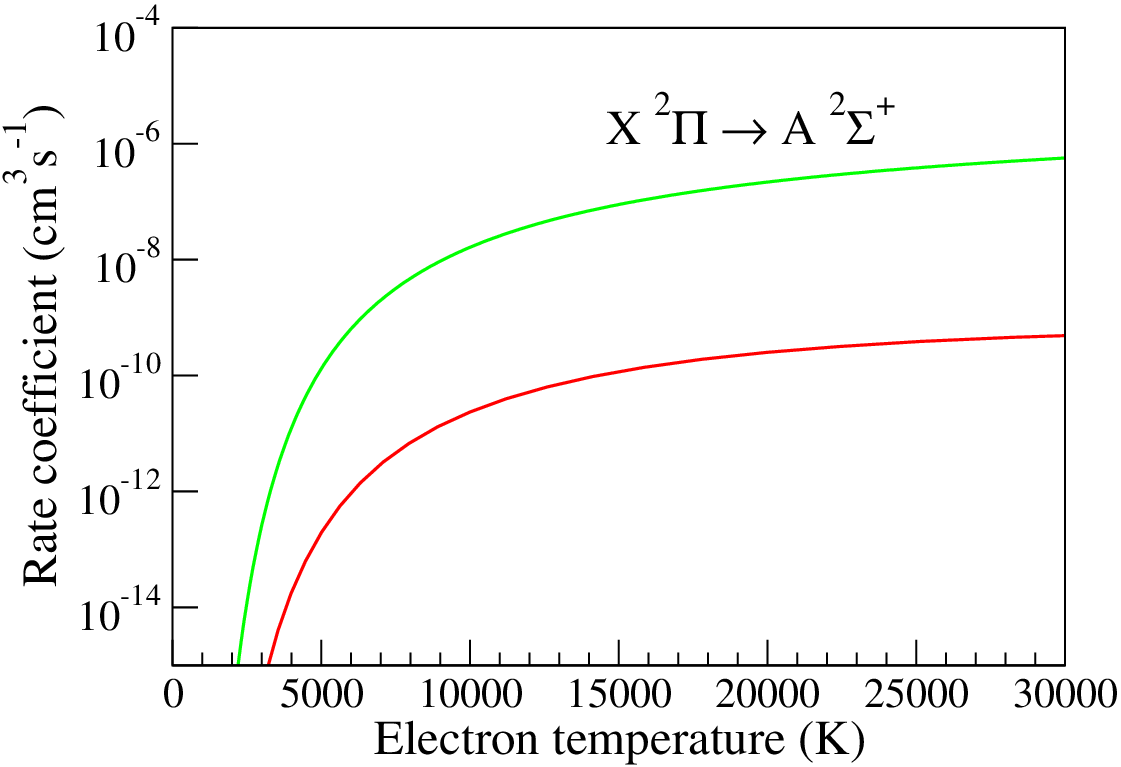}
\end{center}
\caption{Rate coefficients for excitation of the X\,$^2\Pi$ ground
state of the OH molecule to the A\,$^2\Sigma^+$ excited state at 
$R_e=1.8342$ $a_0$. Top curve Riahi \etal Ref. \cite{Riahi06}; bottom 
curve present results. 
\label{Fig:Rate}}
\end{figure}

Figure \ref{Fig:Excitation} shows cross sections for excitation of 
the OH(X\,$^2\Pi$) ground state to the A\,$^2\Sigma^+$, 
a\,$^4\Sigma^-$ and the $1\,^2\Sigma^-$ states. The cross sections 
show sharp features due to negative ion resonances. Figure 
\ref{Fig:Rate} shows the rate coefficients for the $\mathrm{X}\,^2\Pi \rightarrow 
\mathrm{A}\,^2\Sigma^+$. These are compared with the only available rate
coefficients given by Riahi \etal \cite{Riahi06}. The rate 
coefficients of Riahi \etal are obtained by fitting to an Arrhenius
form $k = a\, \theta^b\, \exp{(-c/\theta)}$. Though the shape of the 
rate coefficient curves agree well, the rates given by Riahi \etal 
are much larger than ours, particularly at the higher end of the
temperature range. In terms of electron energy, the rates we presented
are in the range $0 - 3$ eV. In this low energy regime, the R-matrix
method is known to give good results and hence we believe our rates
to be more reliable than those of Riahi \etal.

\begin{figure}[h!]
\begin{center}
\includegraphics[scale=0.4,viewport=0 0 529 375, clip=true]{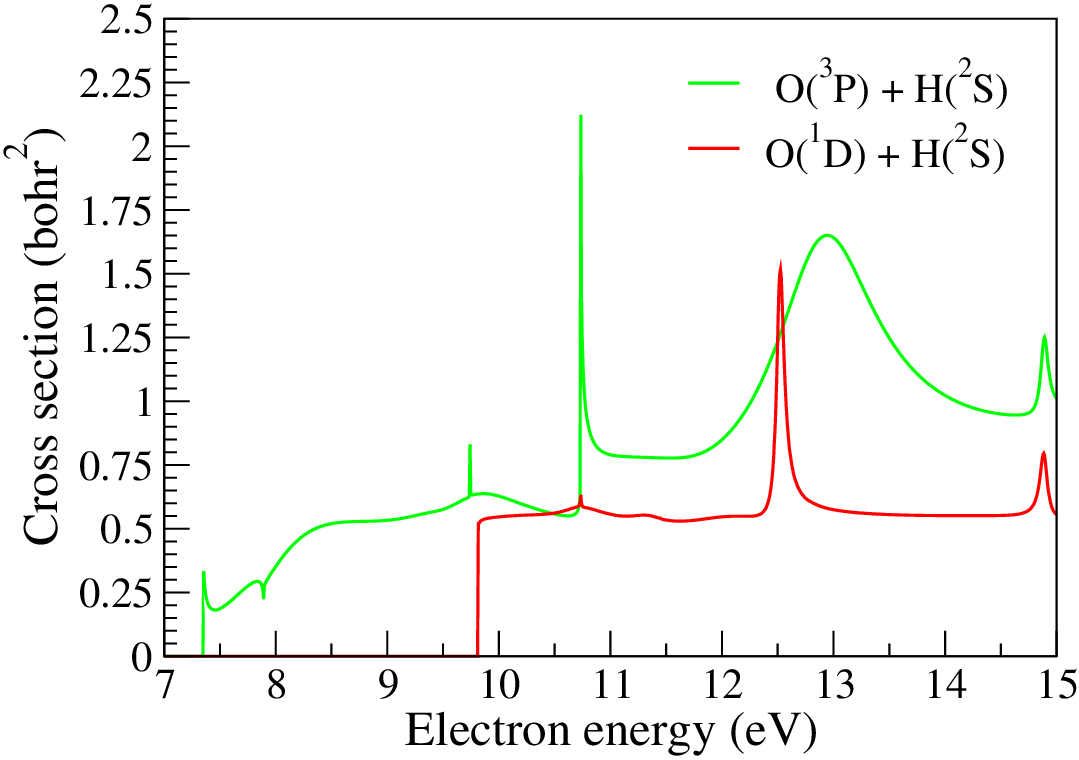}
\end{center}
\caption{Cross section for electron impact dissociation of OH at
$R_e=1.8342$ $a_0$ to the lowest  O($^3$P) + H($^2$S) and
to the excited O($^1$D) + H($^2$S) dissociation channels.
\label{Fig:Dissociation}}
\end{figure}

It is known that electron impact dissociation occurs \emph{via} electronic
excitation, particularly through excitation to dissociative states. 
OH has three repulsive dissociative states, namely the a\,$^4\Sigma^-$, 
$1\,^2\Sigma^-$ and the $b\,^4\Pi$ states, which go to the 
$\mathrm{O}(^3\mathrm{P})+\mathrm{H}(^2\mathrm{S})$ asymptotic limit.
Assuming that excitation to these states above the dissociation threshold ($D_0 = 3.88$~eV)
results in dissociation, 
we present in Fig.~\ref{Fig:Dissociation} our estimate of the cross
section for electron impact dissociation of OH. We predict that
this process  produces a significant quantity of excited,
O($^1$D) atoms. There appears to be
no experimental or theoretical data on this process to compare with.

\subsubsection{Dissociative electron attachment}
To the best of our knowledge, the dissociative electron attachment (DEA)
to OH has never been studied before, theoretically or experimentally, 
though several works on DEA of H$_2$O exist\cite{Compton67,Melton72} (see
also \cite{Itikawa05}). A detailed study of the DEA
process would require full resonance curves and resonance widths across
all internuclear distances considered and is not attempted here. However,
we try to give an estimate of the DEA cross sections using an approximation
developed by Munro \etal \cite{Munro12}

Since the details of the method can be found in Ref. \cite{Munro12},
we only present the essentials. The inputs for the method are the 
resonance positions and widths for OH$^-$ resonances at a single 
geometry, here the equilibrium geometry $R_e=1.8342$ \ao, the PEC 
of the target X\,$^2\Pi$ ground state, and the electron affinity of 
OH, which is taken to be 1.56 eV from column 2 of Table \ref{tab:ea}. 
The target PEC is chosen to be a Morse form with dissociation energy 
$D_e=4.51$ eV. The A\,$^1\Pi$ and a\,$^3\Pi$ resonance potentials were 
chosen to be of a Morse form while the $1\,^3\Sigma^+$ resonance 
potential was chosen to be of an exponentially decreasing form 
following the shapes given by Srivastava and Satyamurthy 
\cite{Srivastava14} and Vamhindi and Nsangou \cite{Vamhindi16} for 
these resonant states. We mention however, that both these works 
\cite{Srivastava14,Vamhindi16} treat resonant part of the A\,$^1\Pi$, 
a\,$^3\Pi$ and $1\,^3\Sigma^+$ potentials like bound states in their 
quantum chemistry calculation and hence must be treated with caution
\cite{jt241}. Moreover, because of their treatment of resonant states as bound, they 
are not able to provide the resonance widths. 

\begin{figure}[h!]
\begin{center}
\includegraphics[scale=0.65,viewport=0 0 421 308, clip=true]{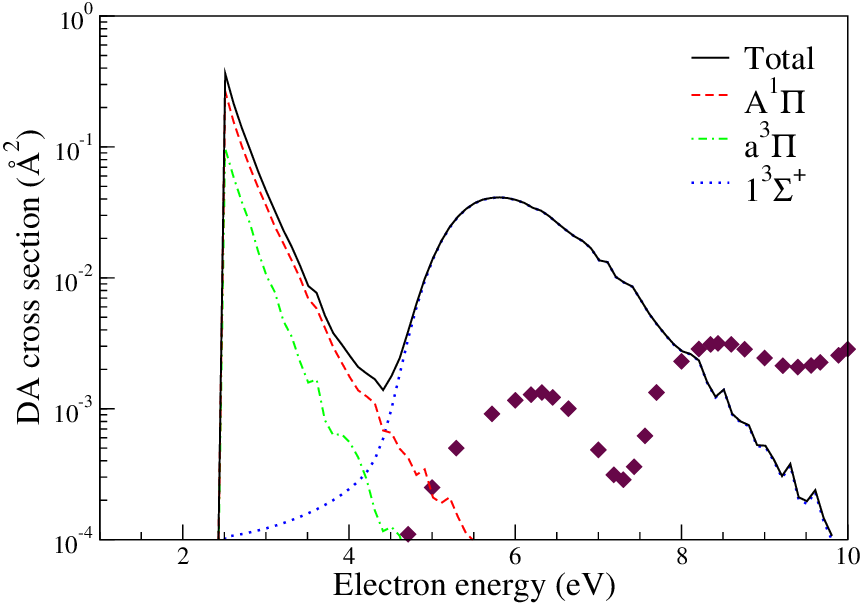}
\end{center}
\caption{Cross section for dissociative electron attachment to OH
for the production of O$^-$ ions. Contributions from different resonant
states are shown in dashed lines with the following convention: blue - 
$1~^3\Sigma^+$, green a~$^3\Pi$, orange A~$^1\Pi$. Filled diamonds: 
experimental values for O$^-$ ion production for dissociative attachment
to H$_2$O from Table 14 of Ref. \cite{Itikawa05}.
\label{fig:dea}}
\end{figure}

Figure \ref{fig:dea} shows our estimate of the DEA cross section. We assume
that DEA produces only O$^-$ since the singlet resonances cannot correlate
with the H$^-$+O($^3$P) limit and all three resonance curves go asymptotically
to  O$^-$+H($^2$S) \cite{Srivastava14}.
Since this is an approximate calculation, 
we are not able to comment on the details. We find a sharp peak 
around 3 eV which is clearly due to the A\,$^1\Pi$, a\,$^3\Pi$ 
resonances and a second broader peak around 6 eV which can be 
attributed to the $1\,^3\Sigma^+$. Since no other results are available 
for comparison, we have shown the experimental values 
\cite{Melton72,Itikawa05} for O$^-$ ion production from the dissociative
electron attachment to H$_2$O which also displays a similar two 
peak structure. We conclude therefore that the DEA process for the
production of O$^-$ ions proceeds via the A\,$^1\Pi$, a\,$^3\Pi$ 
resonances below 5 eV, but above 5 eV the $1\,^3\Sigma^+$ resonance 
mainly drives the DEA process. These cross sections are not insignificant
and should be included in models of OH plasmas.

\section{Conclusion}
\label{sec:conc}
We  study electron collision with OH using the R-matrix
method. Scattering calculations are performed at a single geometry, namely
the OH equilibrium geometry $R_e=1.8342$ \ao to obtain cross sections for
elastic scattering and electronic excitations to the lowest three excited
states, namely the A\,$^2\Sigma^+$, a\,$^4\Sigma^-$ and the $1\,^2\Sigma^-$
states. We also obtained an estimate of the electron impact dissociation 
cross section of OH on the assumption that electronic excitation to the 
states going to the O($^3$P)+H($^2$S) limit above the dissociation
threshold leads to dissociation. The scattering calculations also yield 
OH$^-$ bound states and negative ion resonances in the $e$--OH system.
The X\,$^1\Sigma^+$ ground state of OH$^-$ was found to be bound and 
several resonances that are likely to be important for DEA were 
identified. Since a detailed study of the DEA process is beyond the
scope of this paper, an estimate of the DEA cross section was obtained
using the A\,$^1\Pi$, a\,$^3\Pi$ and  1\,$^3\Sigma^+$ resonances and their
widths at a single geometry. To the best of our knowledge, these cross
sections are being reported for the first time.

A spreadsheet containing our cross sections is provided as supplementary
data to this article.

\section*{Acknowledgement}
VL would to thank Prof. M. Panesi for the kind hospitality at Aerospace Engineering Department at the University of Illinois
at Urbana-Champaign (US) where this work was  completed.

\section*{ORCID iDs}
K Chakrabarti https://orcid.org/0000-0003-0013-5610\\
V. Laporta https://orcid.org/0000-0003-4251-407X\\
J. Tennyson https://orcid.org/0000-0002-4994-5238\\

\section*{References}

\providecommand{\newblock}{}

\end{document}